\documentclass[conference]{IEEEtran}

\ifCLASSINFOpdf

\else

\fi

\usepackage[switch]{lineno}

\begin{document}

\title{Tesla's Autopilot: Ethics and Tragedy}

\author{\IEEEauthorblockN{Aravinda Jatavallabha}
\IEEEauthorblockA{Department of Computer Science\\
North Carolina State University\\
Raleigh, NC-27695\\
arjatava@ncsu.edu}}

\maketitle

\begin{abstract}
The case study delves into the ethical ramifications of an incident involving Tesla's Autopilot, emphasizing Tesla Motors' moral responsibility. Using a seven-step ethical decision-making process, it examines user behavior, system constraints, and regulatory implications. This incident prompts a broader evaluation of ethical challenges in the automotive industry's adoption of autonomous technologies, urging a reconsideration of industry norms and legal frameworks. The analysis offers a succinct exploration of ethical considerations in evolving technological landscapes.
\end{abstract}

\IEEEpeerreviewmaketitle

\section{Introduction}\label{sec:Introduction}
Millions of annual U.S. car accidents result mainly from human error, causing 35,000 fatalities and over \$871 billion in damages \cite{c1}. To address this, Tesla pioneered autonomous vehicle systems like "Autopilot" to eliminate human mistakes and the need for operators in their cars. The rapid advancement of autonomous driving technology is reshaping the transportation landscape, offering increased ease and security but presenting moral dilemmas, exemplified by the fatal incident between Joshua Brown and Tesla's Autopilot system \cite{c2}\cite{c3}. Mr. Brown's sudden death on May 7, 2016, when his Tesla Model S collided with a tractor-trailer in Autopilot mode, underscores the ethical concerns associated with new technological frontiers, mainly when human lives are involved.\par

Tesla reported the first recorded death in over 130 million Autopilot miles, a remarkable contrast to the approximately 94 million miles between fatalities on U.S. roads \cite{c4}. The National Highway Traffic Safety Administration (NHTSA) initiated an assessment, emphasizing the need to determine if the technology performed as expected. Tesla promptly notified the NHTSA, detailing a unique incident on a divided highway where a tractor-trailer crossed the road, resulting in the Model S passing under the trailer and impacting the windshield.\par

The accident occurred as Joshua Brown's Model S traveled east on US-27A, a divided highway in northern Florida. A tractor-trailer, traveling in the opposite direction on the highway, turned left in front of the Tesla. The Tesla was in Autopilot mode. According to Tesla Motors, ``Neither Autopilot nor the driver noticed the white side of the tractor trailer against a brightly lit sky, so the brake was not applied" \cite{c3}. The trailer was high enough off the ground that the car continued under it, shearing off its roof. The car drove off the road and struck two fences and a power pole before stopping.\par

The core issue in Joshua Brown's fatal collision with Tesla's Autopilot system involves the ethical implications of autonomous driving technology. Tesla Motors' ethical responsibilities in implementing and advertising a system requiring constant human supervision, despite being labeled Autopilot, raise concerns about user awareness, safety precautions, and the moral shift from driver assistance to partial automation. Non-ethical aspects, such as technological and legal restrictions, add complexity, necessitating a careful analysis of the Autopilot system's operation and regulatory frameworks. This essay employs a seven-step ethical decision-making framework \cite{c5} to thoroughly examine Tesla Motors' moral responsibilities as a leader in autonomous driving technology. Beyond case details, the analysis underscores the significant consequences for the automotive industry, emphasizing the importance of aligning moral principles and legal systems with advancing technology in a future where humans and machines share the road.

\section{Ethical Decision-Making Framework}
Navigating the moral complexities of Tesla's Autopilot system requires meticulous attention. This section employs a structured seven-step process, used in this study, to analyze the moral implications of autonomous driving technology, covered in the \ref{sec:Introduction}{ Introduction}, \ref{sec:facts}{ Gathering and assessing relevant facts}, \ref{sec:stakeholders}{ Identifying stakeholders}, \ref{sec:options}{ Developing a list of five options}, \ref{sec:testing options}{ Testing options}, and combining the 6th and 7th steps, i.e, tentative and final choices to tackle the problem, in the \ref{sec:conclusion}{ Conclusion}. Following Joshua Brown's tragic incident, each step in this framework serves as a tool to illuminate Tesla Motors' moral responsibility, providing insights beyond the case's specifics for guiding moral decision-making in the dynamic field of technological innovation.

\subsection{Gathering and Accessing Relevant Facts}\label{sec:facts}
Several academic investigations have explored the ethical aspects of the development of autonomous vehicles, with a primary emphasis on imbuing these systems with human ethical standards \cite{c6}\cite{c7}. These studies compare ethical conundrums people face and possible obstacles to widely adopted self-driving automobiles. Nonetheless, there is a crucial void in these conversations since they frequently need to focus more on in-depth analyses of the design process and how people incorporate their moral principles into technology, impacting how society interacts with and uses it.\par

One particular study, ``The Ethics of Crashes with Self-Driving Cars: A Roadmap," \cite{c8} delves into the initial stages of autonomous vehicle testing by prominent tech firms such as Google, Uber, and Tesla. The author draws attention to the different perspectives these corporations hold about the degree of accountability placed on them and how cars should handle moral conundrums when they are about to happen. According to the author, these contradictions indicate a need for more well-defined ethical protocols and preparedness for self-driving cars. Specifically, the author highlights the lack of standard operating guidelines for vehicle algorithms in crash situations with potentially fatal consequences. When discussing accident algorithms for self-driving cars, the author cautions against ignoring essential moral and legal accountability issues.\par

Another study, "Self-Driving Vehicles - an Ethical Overview," \cite{c9}, addresses the potential responsibility gap in a road system with autonomous vehicles. The authors argue that assigning blame solely to the car owner, who may not control the autonomous vehicle, becomes challenging. Accountability is not confined to manufacturer guidelines or the car's intelligence system. Determining fault in a car accident becomes complex due to varying circumstances and damages. The paper emphasizes how societal decisions significantly impact ethical consequences, stressing the need for a systematic approach to ethics in self-driving car technology. \par

While these academic studies offer insightful analysis of potential moral conundrums about AI and autonomous driving, there still needs to be a critical vacuum in understanding how ethical considerations should be incorporated into the design.\par

\subsection{Identifying Stakeholders}\label{sec:stakeholders}
Recognizing the broad spectrum of individuals and entities impacted by decisions is crucial to identifying Autopilot technology stakeholders at Tesla. Primary stakeholders include \textit{Tesla Motors}, the technology creator and producer, with a significant stake in ethical considerations spanning product design, marketing, and user instructions. \textit{Investors and Shareholders} are vital stakeholders, as moral choices with Autopilot impact the company's reputation and financial performance.\par

\textit{Autopilot Users} are essential stakeholders, as the system directly affects their safety. The public and a larger customer base are involved due to the influence on social norms, public trust, and road safety. \textit{Legislators and Regulatory Agencies} also hold significant stakes in creating laws governing autonomous vehicle use.\par

\textit{Rivals} in the autonomous driving space, such as automakers and tech firms, are also stakeholders influenced by Tesla's choices.\par

Specific to this case study, other impactful stakeholders include Joshua Brown's family, victims of Autopilot-related incidents, providing a human face to ethical issues. \par

In the social context, advocacy groups and organizations prioritizing consumer rights, technology ethics, and road safety support moral behavior in autonomous driving.\par
Lastly, as road users alongside autonomous cars, \textit{Pedestrians} are vital stakeholders affected by potential issues or malfunctions.\par

\subsection{List of Options}\label{sec:options}
The following options provide Tesla Motors with various approaches to consider in addressing the moral dilemmas raised by Autopilot. Each choice has its own benefits and drawbacks, and combining these strategies could thoroughly address the intricate ethical situation. \par

\subsubsection{\textbf{Improve User Education and Awareness}}

Provide thorough training resources to guarantee that users of Autopilot are aware of all the features and restrictions of the system. Also, before turning on Autopilot, ensure users complete required tests or training sessions. Stress the value of staying attentive and involved while the system is in operation. Numerous studies also indicate that one of the main causes of crashes is a lack of user education \cite{c10}\cite{c11}.

\subsubsection{\textbf{Limit Autopilot Usage to Specific Driving Situations}}

Implement software updates that limit the usage of Autopilot to specific driving situations, like only using it on highways or clearly defined roads. To ensure safer usage in challenging or dangerous areas, geofencing technology creates virtual limits where Autopilot functionalities are automatically disabled.

\subsubsection{\textbf{Install More Advanced Driver Monitoring Systems}}

Many Tesla owners, as per interviews \cite{c12}, admitted to growing complacent when the Autopilot system is active—a potentially lethal oversight with consequences for both individuals and the environment. To tackle this, one can upgrade the driver monitoring system for Autopilot to guarantee sustained driver focus. Employ advanced technologies like facial recognition and precision steering wheel sensors to gauge driver attention accurately. Implement a high-resolution camera system to capture periodic driver snapshots. Process these images through a specialized neural network model trained to classify varying levels of attentiveness. The neural network's output will be the foundation for a responsive warning and alarm system. Upon detecting signs of reduced attentiveness, the system will trigger real-time alerts, ensuring immediate driver notification. This technical solution integrates cutting-edge facial recognition, sensor technology, and neural network decision-making for precise and timely responses. This enhanced proposal delivers a technically sophisticated approach to driver monitoring, reinforcing Autopilot safety mechanisms with improved accuracy and responsiveness.

\subsubsection{\textbf{Work together with Regulatory and Industry Organizations}}

Establish industry-wide standards for autonomous driving technologies by working cooperatively with other automakers, tech firms, and government agencies. Participate actively in forums and projects to establish moral standards and best practices for creating and implementing autonomous systems.

\subsubsection{\textbf{Constant Improvement via User Input}}

 Provide Autopilot users a reliable feedback channel to share their experiences and comment on how the system works. Utilize user feedback as the basis for ongoing development, modifying the Autopilot system to solve new moral dilemmas and increase security.

\subsection{Testing Options}\label{sec:testing options}
\subsubsection{Publicity Test}

\begin{itemize}
    \item \textbf{Option 1:} Promoting user education demonstrates a commitment to safety and moral technology use, generating positive publicity. Noble, but may elicit different enthusiasm than cutting-edge technology alternatives \cite{c10}.

    \item \textbf{Option 2:} Making usage limitations public demonstrates a commitment to user safety over widespread adoption, generating positive press. Aligns with safety concerns but may face criticism for underutilizing Autopilot capabilities \cite{c13}.

    \item \textbf{Option 3:} Announcing advanced monitoring systems improves public perception, showcasing a proactive security approach. Garners positive press but may raise privacy concerns, resulting in conflicting public responses \cite{c14}.

    \item \textbf{Option 4:} Publicizing partnership activities projects Tesla as a cooperative and responsible industry member, likely well-received by the media.

    \item \textbf{Option 5:} Commitment to using user feedback portrays Tesla as user-focused and constantly evolving. Aligned with a customer-centric strategy, likely receiving positive press. However, poor feedback integration or severe system problems may expose Tesla to criticism \cite{c15}.
\end{itemize}

\subsubsection{Harm Test}

\begin{itemize}
    \item \textbf{Option 1:} Reduces harm by enhancing user understanding and minimizing misuse through comprehensive training, directly tackling potential harm causes. Thorough education significantly diminishes the risk of accidents.

    \item \textbf{Option 2:} Addresses potential danger by limiting Autopilot to safer driving situations, reducing harm and lowering the chance of accidents. However, criticism may arise for limiting system capabilities.

    \item \textbf{Option 3:} Mitigates harm by ensuring driver attentiveness, reducing risks associated with distracted driving. However, privacy concerns and monitoring system effectiveness may vary.

    \item \textbf{Option 4:} Lowers harm by contributing to industry-wide guidelines for responsible, autonomous driving. Focuses on creating a standardized framework, reducing harm from inconsistent regulations. Progress in regulatory discussions may take time.

    \item \textbf{Option 5:} Addresses potential harm by applying user feedback, enhancing system security, and proactively tackling emerging ethical challenges. Effectiveness depends on swift feedback implementation.
\end{itemize}

\subsubsection{Defensibility Test}

\begin{itemize}
    \item \textbf{Option 1:} Easily defensible for focusing on preventative education, lowering accident risk, and emphasizing dedication to user safety. Positions Tesla as a responsible company committed to proactive safety measures \cite{c10}

    \item \textbf{Option 2:} Defensible by emphasizing usage limits for improved safety, prioritizing responsible technology application. However, it might face opposition from users seeking more autonomy.

    \item \textbf{Option 3:} Defensible by highlighting enhanced monitoring's significance in averting mishaps, aligning with the dedication to driver safety. Challenges may arise from privacy concerns.

    \item \textbf{Option 4:} Easily defensible by emphasizing industry-wide norms and proactive participation in ethical issues. Positions Tesla as a responsible corporate citizen contributing to establishing industry standards.

    \item \textbf{Option 5:} Defensible by highlighting the importance of user input in enhancing system security and addressing moral issues. Effectiveness depends on the robustness of the feedback system and the swift implementation of necessary changes.
\end{itemize}

\subsubsection{Reversibility Test}
\begin{itemize}
    \item \textbf{Option 1:} This test is irrelevant for this option as everyone, including me, would advocate spreading more awareness and education to users as it will only benefit them.

    \item \textbf{Option 2:} Many, myself included, would likely adhere to this option, acknowledging the risks associated with using Autopilot in inappropriate areas where unforeseen issues arise.

    \item \textbf{Option 3:} Many individuals, myself included, would support Tesla's integration of this option into their Autopilot system, recognizing its benefits. However, Tesla must establish transparent privacy policies and inform users about their rights, addressing the associated privacy concerns in implementing this option.

    \item \textbf{Option 4:} The test is irrelevant because increased collaboration between the company and regulatory bodies is universally endorsed, including my own perspective, as it aligns with ethical and responsible practices in developing autonomous driving technology.

    \item \textbf{Option 5:} This option, focused on continuous improvement, is likely accepted by myself and others. Success depends on an efficient feedback system and prompt implementation, but rejection may occur if these aspects face challenges.
\end{itemize}

\subsubsection{Virtue Test}
\begin{itemize}
    \item \textbf{Option 1:} Advocacy for responsible tech use, demonstrating a commitment to user safety and ethical Autopilot use. Aligns with virtues like integrity, contributing to a digital environment prioritizing safety.

    \item \textbf{Option 2:} Champions caution and safety in autonomous driving, reflecting prudence and responsibility. Prioritizes safety over convenience, dedicating to thoughtful choices and risk reduction.

    \item \textbf{Option 3:} Propels toward driver safety and attentiveness, reflecting diligence and dedication. Showcases ongoing efforts to improve safety, aligning with virtues such as responsibility and continuous improvement.

    \item \textbf{Option 4:} Reflects collaboration and advocacy for ethical industry standards. I would commit to working with regulatory bodies that align with virtues like cooperation and integrity, contributing to moral business practices.

    \item \textbf{Option 5:} Advocates for user-centric and responsive tech development. Commitment to incorporating user feedback aligns with virtues like responsiveness and dedication to user satisfaction. Actions contribute to ongoing efforts for ethical and user-focused advancement.
\end{itemize}

\subsubsection{Colleague Test}
\begin{itemize}
    \item \textbf{Option 1:} Generally favored by colleagues, demonstrating dedication to user education and safety, aligning with shared principles of responsible tech usage. Some may argue that user education alone is insufficient; technical limitations should be addressed concurrently.

    \item \textbf{Option 2:} Mainly endorsed by colleagues, aligning with shared ideals of putting safety before widespread use. Some may express concerns about limitations on user convenience and the potential negative impacts of adopting autonomous technology.

    \item \textbf{Option 3:} Favored by colleagues, consistent with shared principles of responsibly putting safety first and developing technology. Some may have privacy concerns about advanced driver monitoring, leading to partial disapproval, but education can address these issues.

    \item \textbf{Option 4:} Endorsed by colleagues, consistent with mutual values of business cooperation and ethical standards development. Some skepticism among a few colleagues about collaboration effectiveness with regulatory bodies may lead to reservations.

    \item \textbf{Option 5:} Favored by colleagues, consistent with shared principles of continuous improvement and user-centric tech development. Some concerns about the practicality of incorporating user feedback may be raised, leading to partial disapproval.
\end{itemize}

\subsubsection{Professional Test}
\begin{itemize}
    \item \textbf{Option 1:} Will be approved by the ethics committee for aligning with industry expectations, prioritizing user education for ethical autonomous features. Educating users will be seen as a responsible tech usage approach.

    \item \textbf{Option 2:} Will be positively acknowledged by the ethics committee for conforming to industry standards. Restricting Autopilot for enhanced safety will be considered a responsible and ethical decision, although the committee might advocate making the privacy risks involved more transparent.

    \item \textbf{Option 3:} Will gain approval for aligning with industry norms, prioritizing driver safety through enhanced monitoring. Some backlash on privacy issues might be heard.

    \item \textbf{Option 4:} Will be strongly supported by the ethics committee for aligning with ethical standards in collaboration with regulatory bodies. Developing industry-wide guidelines will be viewed as a responsible and ethical contribution.

    \item \textbf{Option 5:} Will receive cautious approval from the ethics committee. While valuing user feedback, concerns about relying solely on it for ongoing improvement may exist. Balancing user-centricity and technical expertise will be essential for ethical tech advancement.
\end{itemize}

\subsubsection{Organizational Test}
\begin{itemize}
    \item \textbf{Option 1:} Aligns with Tesla's user safety commitment, likely gaining positive feedback from the ethics officer. Emphasizing user education reflects a proactive safety approach.

    \item \textbf{Option 2:} Aligns with Tesla's safety commitment, positively reviewed by the ethics officer. Focusing on limiting Autopilot aligns with principles and guidelines.

    \item \textbf{Option 3:} Demonstrates dedication to safety tech, likely positively received by the ethics officer. Installing advanced monitoring reflects a commitment to innovation and ethical driving practices.

    \item \textbf{Option 4:} Highlights Tesla's commitment to moral behavior and industry advancement. This option will likely receive positive feedback, showcasing dedication to ethical practices and collaboration.

    \item \textbf{Option 5:} Demonstrates a proactive approach to resolving ethical problems, aligning with Tesla's values. This option will likely be positively reviewed, reflecting a commitment to continuous improvement and user-centric tech development.
\end{itemize}

\section{Technical Aspects of Tesla's Autopilot}\label{sec:Technical}

Tesla's Autopilot relies heavily on a combination of computer vision, sensor fusion, and neural networks to interpret and respond to the driving environment. Key technologies include deep learning models for object detection, lane detection, and decision-making, all of which interact with hardware components such as cameras, radar, ultrasonic sensors, and LiDAR.

\subsection{Neural Networks in Autopilot}

The core of Tesla's Autopilot system is a multi-layer convolutional neural network (CNN) trained on vast amounts of driving data. The CNN is responsible for identifying objects on the road, such as vehicles, pedestrians, and road signs. The system also uses a recurrent neural network (RNN) for temporal data processing, analyzing sequences of frames to predict movement and trajectory.

These models, trained on diverse driving conditions, form the basis of Autopilot’s decision-making process. The neural networks are fed real-time data from cameras and other sensors, which helps the vehicle identify objects, classify their relevance (e.g., pedestrian vs. vehicle), and make split-second decisions.

\subsection{Limitations of Current Machine Learning Models}

Despite the sophistication of these models, limitations in dataset coverage and edge-case scenarios lead to failures such as the one described in the Joshua Brown case. The system’s inability to recognize the white side of the tractor-trailer against a bright sky, as documented in the incident, highlights the challenge of generalizing from training data. This points to the need for improvements in both model accuracy and real-time sensor data processing, as well as an emphasis on robustness to rare but critical scenarios (i.e., edge cases).

\subsection{Ethical Implications of Training Data}

A critical consideration in the ethics of machine learning is the quality and diversity of the training data. In Tesla's case, while the system is trained on millions of miles of driving data, there is an ethical concern regarding the responsibility of the company to ensure this data encompasses a wide variety of environments, weather conditions, and unusual road circumstances. Failure to do so not only increases the likelihood of accidents but also raises questions about bias in the algorithm—whether certain types of roads or conditions are underrepresented.

\subsection{Sensor Fusion for Enhanced Perception}

Tesla's Autopilot also employs sensor fusion to combine inputs from multiple sensors, including radar and ultrasonic sensors, to make more informed decisions. The fusion of data from these sources is processed through a Kalman filter, which helps the system predict the position of objects in real-time. This, combined with the visual data from cameras, allows for a more comprehensive understanding of the environment. However, technical limitations in sensor fusion algorithms, such as latency in data processing or inaccuracies in spatial mapping, can lead to tragic outcomes, as in the Joshua Brown case.

\subsection{Future Developments and the Role of Reinforcement Learning}

Future iterations of autonomous systems may benefit from reinforcement learning, where the model learns optimal driving behavior through trial and error in a simulated environment. Tesla has started exploring this avenue by using driving simulators to train models to better predict and react to novel situations. Ethical concerns arise here as well, particularly regarding the transparency of how these models are validated before deployment on real roads.

\subsection{Interpretability of AI Models and Ethical Accountability}

One of the critical challenges in autonomous driving is the interpretability of neural networks. In scenarios like the Joshua Brown accident, it is difficult to determine why the model failed to recognize the trailer. Interpretability is crucial for ensuring ethical accountability in AI-driven systems. Current research is focusing on methods like saliency maps and SHAP (SHapley Additive exPlanations) values to provide more transparency in model decisions.

Ethical frameworks such as the one discussed in this paper must also consider the need for more interpretable AI models. By ensuring that the models can provide a reason for each decision, regulators, engineers, and the public can have greater confidence in the safety and reliability of autonomous systems.

\section{Ethical Decision-Making in Autonomous Systems}

With the technical depth of Tesla’s Autopilot understood, we can now revisit the ethical decision-making framework. The ethical implications extend beyond the scope of user education or regulation; they touch upon the very design of the machine learning models and their training data, sensor usage, and interpretability. This highlights the interconnectedness of technical choices with ethical outcomes, reinforcing the need for continuous advancements in both AI technologies and ethical standards.

\section{Conclusion}\label{sec:conclusion}
Addressing the challenges with Tesla's Autopilot system requires both technical and ethical advancements. My tentative choice would be implementing Option 3, which incorporates advanced driver monitoring systems utilizing precise sensors, facial recognition, and a dedicated neural network model. This solution consistently evaluates driver attentiveness, triggering real-time alerts in instances of diminished focus, ensuring an additional layer of safety.\par

However, recognizing that this choice alone may not suffice, especially if users are not adequately educated about the proposed monitoring system, misconceptions and trust issues may arise, potentially leading to privacy concerns. To mitigate this, a proactive public education campaign is crucial. Transparent communication on stringent privacy measures and safety priorities must accompany any technical improvement. A proactive stance in educating the public about their privacy rights and the system’s capabilities is essential to building trust.\par

Furthermore, the combination of Options 1 and 3 offers a robust and balanced approach to reduce the likelihood of similar incidents in the future. Vigilant driver education minimizes the risk of misuse, while advanced monitoring systems provide a technological safety net. Collaborating with industry stakeholders, regulators, and policy-makers will be essential in creating a sustainable framework for autonomous driving. By doing so, Tesla can reinforce its commitment to both innovation and safety.\par

Combining \textbf{technical advances in machine learning}, such as improved object detection, interpretability of AI models, and sensor fusion, with \textbf{ethical principles}, is key to addressing the challenges faced by autonomous driving systems like Tesla's Autopilot. This comprehensive strategy considers not only technological advancements but also human-centric concerns, contributing to the development of a reliable and morally sound autonomous driving system. The collective impact instills confidence among users and stakeholders, reaffirming Tesla's dedication to safety, transparency, and innovation. Therefore, my final recommendation is the implementation of both Options 1 and 3, ensuring a safer future for autonomous driving technology.

\section*{Acknowledgment}
I sincerely thank Dr. Timothy Menzies and the Teaching Assistants for their assistance and dedication during the course. Their valuable contributions and commitment to fostering a conducive learning environment have greatly enriched my academic experience.

\end{document}